\documentclass[manuscript,screen]{acmart}
\AtBeginDocument{%
  \providecommand\BibTeX{{%
    \normalfont B\kern-0.5em{\scshape i\kern-0.25em b}\kern-0.8em\TeX}}}

\setcopyright{rightsretained}
\copyrightyear{2023}
\acmYear{2023}
\acmDOI{}

\acmConference[CCAI 2023]{CHI 2023 Workshop on Child-centred AI Design: Definition, Operation and Considerations}{April 23, 2023}{Hamburg, Germany}
\acmBooktitle{CHI 2023 Workshop on Child-centred AI Design: Definition, Operation and Considerations, April 23, 2023, Hamburg, Germany}

\begin{document}

%%
%% The "title" command has an optional parameter,
%% allowing the author to define a "short title" to be used in page headers.
\title{Participatory Design of AI with Children: Reflections on IDC Design Challenge }

%%
%% The "author" command and its associated commands are used to define
%% the authors and their affiliations.
%% Of note is the shared affiliation of the first two authors, and the
%% "authornote" and "authornotemark" commands
%% used to denote shared contribution to the research.
\author{Zhen Bai}
\email{zbai@cs.rochester.edu}
\orcid{0000-0002-3258-0228}
\affiliation{%
  \institution{Department of Computer Science, University of Rochester}
  \streetaddress{250 Hutchison Rd}
  \city{Rochester}
  \country{United States}
}

\author{Frances Judd}
\affiliation{%
  \institution{Bennett Day School}
    \country{United States}
}
\email{frances.judd@bennettday.org}

\author{Naomi Polinsky}
\affiliation{%
  \institution{Department of Psychology, Northwestern University}
    \country{United States}
}
\email{naomipolinsky2022@u.northwestern.edu}

\author{Elmira Yadollahi}
\orcid{0000-0001-7091-0104}
\affiliation{%
 \institution{Division of Robotics, Perception and Learning, KTH Royal Institute of Technology}
   \country{Sweden}
 }
 \email{elmiray@kth.se}

%%
%% By default, the full list of authors will be used in the page
%% headers. Often, this list is too long, and will overlap
%% other information printed in the page headers. This command allows
%% the author to define a more concise list
%% of authors' names for this purpose.
%\renewcommand{\shortauthors}{Trovato and Tobin, et al.}
%\renewcommand{\shorttitle}{Participatory Design of AI with Children}

%%
%% The abstract is a short summary of the work to be presented in the
%% article.
\begin{abstract}
 Children growing up in the era of Artificial Intelligence (AI) will be most impacted by the technology across their life span. Participatory Design (PD) is widely adopted by the Interaction Design and Children (IDC) community, which empowers children to bring their interests, needs, and creativity to the design process of future technologies. While PD has drawn increasing attention to human-centered AI design, it remains largely untapped in facilitating the design process of AI technologies relevant to children and their communities. In this paper, we report intruiguing children’s design ideas on AI technologies resulting from the “Research and Design Challenge” of the 22nd ACM Interaction Design and Children conference (IDC 2023). The diversity of design problems, AI applications and capabilities revealed by the children’s design ideas shed light on the potential of engaging children in PD activities for future AI technologies and education. We discuss opportunities and challenges for accessible and inclusive PD experiences with children. 
\end{abstract}

%%
%% The code below is generated by the tool at http://dl.acm.org/ccs.cfm.
%% Please copy and paste the code instead of the example below.
%%

\begin{CCSXML}
<ccs2012>
<concept>
<concept_id>10003120.10003123.10010860.10010911</concept_id>
<concept_desc>Human-centered computing~Participatory design</concept_desc>
<concept_significance>500</concept_significance>
</concept>
</ccs2012>
\end{CCSXML}

\ccsdesc[500]{Human-centered computing~Participatory design}

%%
%% Keywords. The author(s) should pick words that accurately describe
%% the work being presented. Separate the keywords with commas.
\keywords{Participatory Design, Child-Centered AI, Interaction Design and Children, AI Literacy}

%% A "teaser" image appears between the author and affiliation
%% information and the body of the document, and typically spans the
%% page.
\begin{teaserfigure}
  \includegraphics[width=\textwidth]{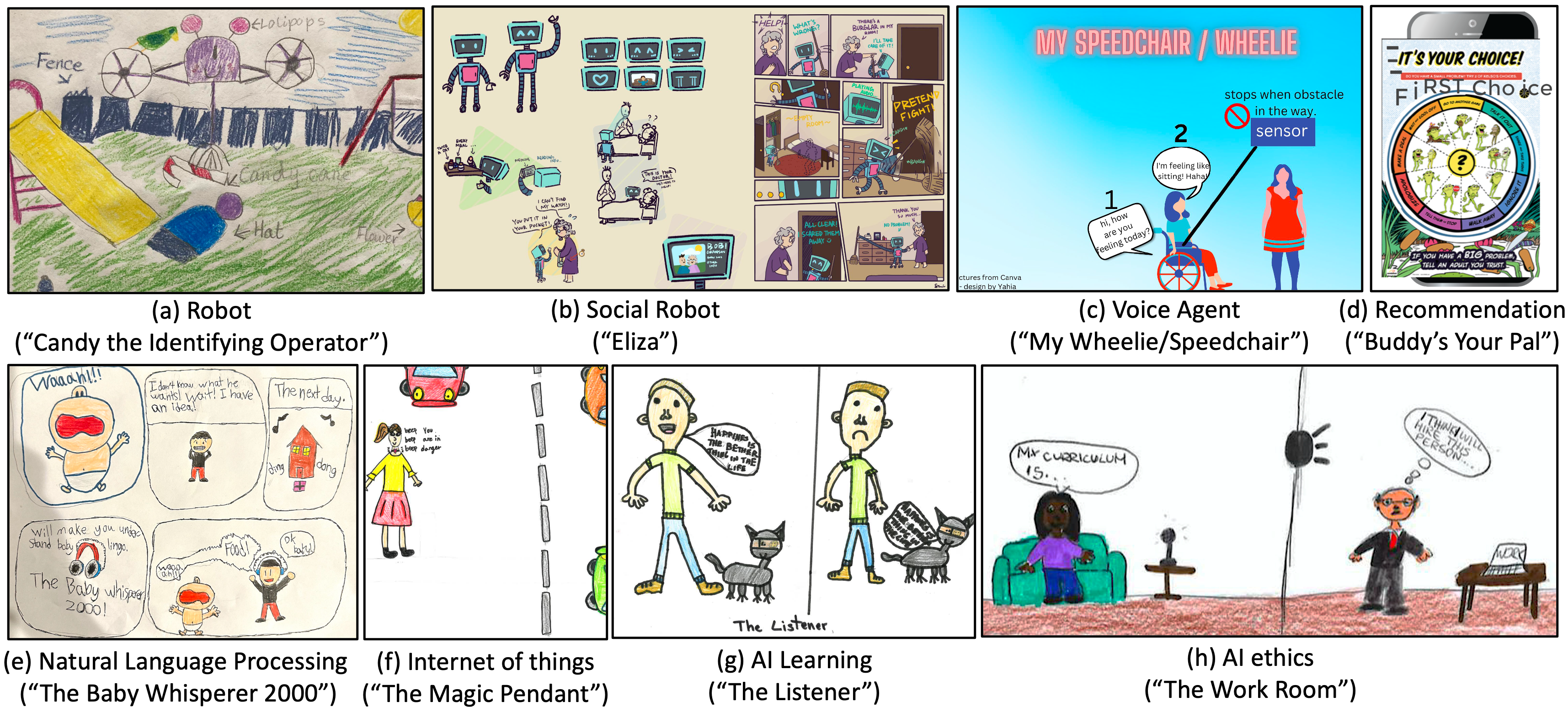}
  \caption{Example design ideas for different AI applications.}
  \Description{Example design ideas for different AI applications.}
  \label{fig:teaser}
\end{teaserfigure}

%\received{20 February 2007}
%\received[revised]{12 March 2009}
%\received[accepted]{5 June 2009}

%%
%% This command processes the author and affiliation and title
%% information and builds the first part of the formatted document.
\maketitle

\section{Introduction}
Participatory design (PD) has been a long-established research method in the Interaction Design and Children (IDC) community \cite{druin1999cooperative,iversen2017child}. It emphasizes active participation of children to bring their interests, needs, and creativity to the center of the design process for future technologies. Children growing up in the Artificial Intelligence (AI) era will be deeply impacted by technological advances in AI, yet PD practices are largely untapped in facilitating the design process of AI technologies relevant to children and their community. This paper reports children’s design ideas on the theme of \textbf{\textit{“Smart Communities: Rebuilding a compassionate world!”}} resulting from the “Research and Design Challenge” of the 22nd ACM Interaction Design and Children Conference (IDC 2023) . The theme presents a bold attempt to learn children’s vision of AI technologies in response to the rapid growth of AI-powered technologies for child users, as well as the emergence of children’s AI education (e.g., AI4K12, AI4ALL) and ethics (e.g., Generation AI).
      
The Challenge is composed of two phases: Phase 1 invites children to create design ideas of future AI technologies in fostering kindness, equality, and sustainability in our communities. Children are encouraged to describe their ideas with drawings, pictures or other forms that best express an initial impression of their ideas. Phase 2 invites UX researchers and designers to submit a design concept that responds to the same theme by iterating on one or more of the ideas the children presented in Phase 1. This study focuses on the child ideation created in Phase 1. 

The two essential questions include: (1) how capable are children to produce authentic design ideas around AI technologies, and (2) what are the opportunities and challenges of child PD in designing AI-powered future technologies?

\section{THE DIVERSITY OF CHILDREN’S DESIGN IDEAS}
We received 60 intriguing design ideas created by children ages 4 to 16 years old (\textit{Mean} = 9.80, \textit{SD} = 1.76). Among these ideas, 60\% were created at K-12 schools, 35\% at home, and 5\% at informal learning spaces. There are five common problems in line with the “Smart Communities” theme: (1) helping people: (a) kids, families, schools (15\%), (b) people with special needs (8\%), (c) people from underserved communities (12\%), and (d) everyone (12\%); (2) protecting the environment (17\%); (3) engendering kindness (27\%); (4) supporting connections (7\%); (5) facilitating equality (3\%). The complete set of child design ideas can be fond at: https://idc.acm.org/2023/research-design-challenge/.

\subsection{AI applications}
Intelligent agent is the most popular AI application (71\%) (e.g., Fig.1(a)-(c)), including non-social robots that can plan and carry out actions (32\%) (e.g., a robot collects lost items on the playground), followed by social robot (22\%) (e.g., a conversational robot that takes care of elderly) and voice agent (17\%) (e.g., a speaking wheelchair to chit chat with people) that can understand and communicate with humans. Other AI applications include screen-based recommendation system (10\%) (e.g., Fig.1(d) a mobile app that helps a child to handle social situations), Natural Language Processing (NLP) involving language translation and generation (5\%) (e.g., Fig.1(e) translate babies’ sounds to their needs), and Smart Internet of Things (IoT) (7\%) (Fig.1(f) a smart pendant to warn people of danger). A small portion of ideas involve limited AI technologies or fictional technologies (8\%) (e.g., a magic bubble surrounding the earth to deliver medicine).

\subsection{AI capabilities}

In terms of AI capabilities based on the “Big Five AI Ideas” for K12 education \cite{touretzky2019envisioning}, three AI capabilities are commonly involved, namely \textbf{\textit{Perception}} (using sensors to understand the world), \textbf{\textit{Representation \& Reasoning}} (reasoning about complex problems based on representations of the world), and \textbf{\textit{Natural Interaction}} (understand and communicate with humans using natural interactions such as speech). Although the design ideas reveal a gap of knowledge in the other two AI capabilities, namely \textbf{\textit{Learning}} (learn from data) and \textbf{\textit{Societal Impact}} (AI ethics), there are sparks of design ideas that show children’s emerging understanding in those areas. For example, the “The Listener” idea (Fig. 1(g)) is a robot pet dog that learns happy words from their humans and says these words to them when they are sad. “The Work Room” idea (Fig.2 (h)) utilizes speech synthesis technologies to support gender equality in the context of employment by hiding gender and ethnicity information from one’s voice.

\section{OPPORTUNITIES AND CHALLENGES FOR AI PD WITH CHILDREN}
We identify opportunities in three aspects: (1) \textbf{\textit{Envisioning future AI technologies}}: the diverse range of authentic problems, AI applications, and capabilities revealed in the children’s design ideas demonstrate a vast potential for children to provide critical design inspirations by serving as informants of their own needs and preferences from personal experiences, also serving as designers to create low-fidelity prototypes that represent their design ideas; (2) \textbf{\textit{Promoting children’s AI literacy}}: the fictional design of AI technologies may empower children in shaping the development of future AI technologies, and enhance critical understanding and reflection of AI that impact their everyday lives \cite{iversen2017child}. The PD activities may also promote AI literacy through a strong sense of authenticity and real-world applicability as key motivations for computational thinking \cite{weintrop2016defining}; (3) \textbf{\textit{Informing K-12 AI educators}}: The design ideas reveal specific gaps and misconceptions in AI literacy such as “learning from data” and “recognizing AI” (telling whether or not AI is involved in an application) \cite{long2020ai}. The high percentage of intelligent agents appearing among the design ideas also corroborates with existing knowledge that children tend to anthropomorphize intelligent systems \cite{heinze2010action, druga2017hey}. Such tendencies may lead to rich learning opportunities for children to develop critical thinking towards trustworthiness of AI systems \cite{long2020ai}.

We also identify two main challenges: (1) \textbf{\textit{Limited technical knowledge in AI}}: Existing PD practices with children often focus on familiar interfaces such as game, mobile, and digital media \cite{yip2013brownies,bonsignore2013embedding}, but children are less familiar with AI technologies due to limited access to AI literacy and technical exposure. Thanks to recent educational initiatives such as AI4K12 and AI4ALL, several learning environments have been created to expose K-12 students to fundamental AI areas (\cite{touretzky2019envisioning} for a review). Nevertheless, many children, especially those from low and medium socio-economic schools, may still struggle to advance their AI understanding mainly due to lack of prior experience in computing \cite{druga2019inclusive}. There is emerging research in K-12 AI education that integrates fictional design with learning AI concepts, programming, and ethics \cite{druga2018growing, ali2019constructionism}, which shed light on a tandem approach of AI PD and AI literacy with children. (2) \textbf{\textit{Limited adult support for AI PD with children}}: PD with children relies heavily on adult facilitation. Research also shows that parent involvement is critical in supporting children’s AI learning experience \cite{beals2006robotic, freed2012fluffy, long2020ai}. As an initial attempt, the Research and Design Challenge demonstrated a viable set of prompts to engage children in fictional design of future AI with the assistance of teachers and parents, both with limited AI knowledge. These prompts include a concise introduction of common AI capabilities and technologies relatable to children’s everyday experiences (e.g., movie recommendation, mood detection, and self-driving cars), and a series of questions that aim to elicit children’s empathy in thinking of AI-powered solutions to address meaningful problems centered with fostering compassion within communities such as treating each other kindly and equally and caring for the sustainability of our environment. We call for community efforts to provide new toolkits and resources to enable accessible and inclusive AI PD experiences through child-adult partnerships, with the overarching goal to advance digital citizenship for children and members of their communities with diverse knowledge, abilities, and socio-cultural backgrounds in the era of AI.

\bibliographystyle{ACM-Reference-Format}
\bibliography{Manuscript}

\end{document}